\documentstyle[prl,aps,epsf]{revtex}         
\begin{document}

\preprint{Preprint}
\draft{}
\twocolumn[\hsize\textwidth\columnwidth\hsize
           \csname @twocolumnfalse\endcsname

\title{Observation of Two-dimensional Spin Fluctuations in the Bilayer
Ruthenate Sr$_3$Ru$_2$O$_7$ by Inelastic Neutron Scattering}

\author{L.~Capogna$^{1,*}$,
E.M.~Forgan$^{1,\dagger}$,
S.M.~Hayden$^{2}$,
A.~Wildes$^{3}$,
J.A.~Duffy$^{2,4}$,
A.P.~Mackenzie$^{1,5}$,
R.S.~Perry$^{1,6}$,
S.~Ikeda$^{6,\S}$,
Y.~Maeno$^{6,7}$,
S.P.~Brown$^{1}$.
}

\address{$^{1}$School of Physics and Astronomy, University of Birmingham, Birmingham B15 2TT, U.K. \\
$^{2}$H.H. Wills Physics Laboratory, University of Bristol, Bristol BS8 1TL, U.K. \\
$^{3}$Institut Laue-Langevin, 6 Rue Jules Horowitz, F38042 Grenoble Cedex, France \\
$^{4}$Department of Physics, University of Warwick, Coventry CV4 7AL, U.K. \\
$^{5}$School of Physics and Astronomy, University of St. Andrews, St. Andrews KY16 9SS, U.K. \\
$^{6}$Department of Physics, Kyoto University, Kyoto 606-8502, Japan \\
$^{7}$CREST, Japan Science and Technology Corporation, Kawaguki, Saitama 332-0012, Japan \\}
\date{\today}
\maketitle
\widetext

\begin{abstract}

We report the first observation of two-dimensional incommensurate magnetic fluctuations
in the layered metallic perovskite Sr$_3$Ru$_2$O$_7$. The wavevectors
where the magnetic fluctuations are strongest are different from those
observed in the superconducting single layer ruthenate Sr$_2$RuO$_4$
and appear to be determined by Fermi surface nesting. No antiferromagnetic ordering
is observed for temperatures down to 1.5~K. For temperatures $T \gtrsim$20~K,
the fluctuations become predominately ferromagnetic. Our inelastic neutron scattering
measurements provide concrete evidence of the coexistence of competing
interactions in Sr$_3$Ru$_2$O$_7$ and of the low energy scale of the fluctuations.
\end{abstract}
 ]
\narrowtext

The nature of magnetic correlations in layered oxide perovskites such
as cuprates, manganites and ruthenates is at the heart of
theoretical and experimental challenges in contemporary solid state
physics. In recent years, the discovery of unconventional
superconductivity in the single-layer ruthenate
Sr$_2$RuO$_4$~\cite{Maeno214} has generated significant interest in
this and related ruthenates. The experimental observation of low
energy incommensurate 2D spin fluctuations in
Sr$_2$RuO$_4$~\cite{Sidis,Servant} has raised the question
of the relevance of spin
fluctuations to the mechanism producing {\it p}-wave
pairing in this material. The closest relative of Sr$_2$RuO$_4$, the
bilayer Sr$_3$Ru$_2$O$_7$, is a paramagnet where ferromagnetic and
antiferromagnetic correlations may be in competition,
and ferromagnetism can be induced
by pressure or impurities~\cite{Maeno327,Ikeda,Liu,Rob,May,Cao}.
In high-quality single crystals
of Sr$_3$Ru$_2$O$_7$, the resistivity exhibits a Fermi-liquid
$T^2$ temperature dependence below 10~K~\cite{Ikeda,Lucia}. However,
a moderate magnetic field induces a metamagnetic transition,
which is accompanied by a striking deviation from
Fermi liquid behaviour~\cite{Rob,Grigera}.
Sr$_3$Ru$_2$O$_7$ appears to be a strong candidate to exhibit a
metamagnetic quantum critical end-point,
driven by the magnetic field and characterized by the absence of spontaneous
symmetry breaking~\cite{Grigera}. In this letter we report the first
observations of low-energy spin fluctuations in Sr$_3$Ru$_2$O$_7$ as
measured by inelastic neutron scattering.

With respect to the conducting and magnetic properties of
Sr$_3$Ru$_2$O$_7$, the fundamental building blocks of its crystal
structure are the RuO$_2$ bilayers joined by an SrO layer.  These slabs
are separated along the crystal $\bbox{c}$-direction by two rock
salt-type layers of SrO which decouple the slabs electronically and
magnetically. At the same time, and in contrast to the single layer
compound Sr$_2$RuO$_4$, the RuO$_6$ octahedra in Sr$_3$Ru$_2$O$_7$ are
rotated around the $\bbox{c}$-axis, by $\sim7^\circ$, changing the
unit cell from body-centred tetragonal to a $\sqrt{2} \times \sqrt{2}$ larger
face-centred orthorhombic cell~\cite{Shaked}. The rotation is expected to reduce
the in-plane Ru-Ru hopping, and hence increase the density of states
at the Fermi level~\cite{Mazin}, which may
enhance the magnetic fluctuations.

Single crystals of Sr$_3$Ru$_2$O$_7$ were grown using a mirror
furnace, and were checked for homogeneity and purity by magnetic,
resistive, and crystallographic measurements. All crystals used in
this study showed the characteristic peak in susceptibility at a temperature
of 17~K and no ferromagnetism. For the inelastic neutron
scattering experiments, three crystals were mounted so that their axes
coincided to form a mosaic sample with total mass 0.9 g.
The sample was mounted in a cryostat on
the cold neutron three-axis spectrometer IN14 at the ILL. For
simplicity, we describe our results using the undistorted tetragonal
cell of the compound, which has the $a$ and $b$ lattice parameters
equal to the in-plane Ru-O-Ru distance, 3.87~\AA. The
$\bbox{c}$-axis is perpendicular to the RuO$_2$ planes and has
the magnitude 20.7~\AA, which is twice the spacing of the RuO$_2$
bilayers, reflecting the body-centred stacking of bilayers~\cite{Shaked}.
Using this unit cell, the main nuclear Bragg peaks of
the 3-D structure occur at points $(h, k, l)$ in reciprocal space
with integer $h, k$ and $l$ and $(h+k+l)$ even. The less intense
ones, arising from the rotations of the octahedra, occur at some of
the points where $h$ and $k$ are half-integral and $l$ is an integer.

\begin{figure}
\centerline{\epsfysize=12.4cm \epsfbox{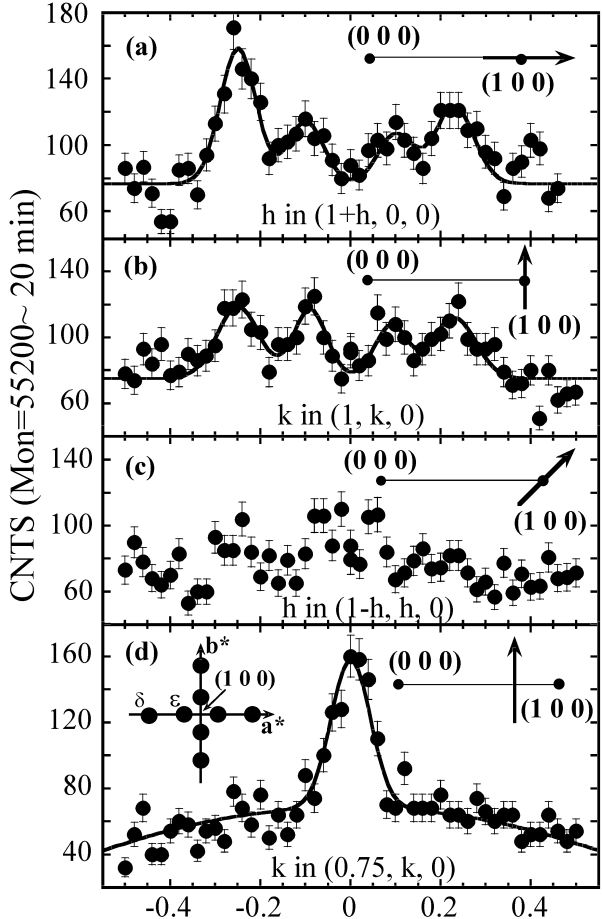}}
\caption{Inelastic neutron scattering at 1.5K: intensity versus
wavevector along lines in the $(h,k,0)$ plane,
at a constant energy transfer of 2 meV.
Solid lines are a sum of Gaussians as a guide to the eye.
The the fwhm $Q$ resolution, calculated for the relaxed collimation of our setup,
is comparable to the width of the symbols.
The inset to (d) indicates schematically the $\bbox{Q}$ positions
where peaks are observed. All measurements in this Letter,
except those in Fig.2,
were taken with a varying incident energy and constant final energy
of 4.97 meV, with a cooled beryllium filter before
the analyser to remove higher order contamination.}
\label{fig:qdep}
\end{figure}

We performed extensive measurements with $(h,k,0)$ as the scattering plane
and further measurements in the $(h,0,l)$ plane. Unlike Sr$_2$RuO$_4$,
magnetic fluctuations at our base temperature of 1.5~K were {\it
not} observed to peak along the $(h,h,0)$ direction from a reciprocal
lattice point; instead they were detected along $(h,0,0)$.
Figure~\ref{fig:qdep} shows representative scans along
major symmetry directions at a constant energy transfer (from neutrons
to the sample) of 2~meV.
Figure~\ref{fig:qdep}(a) shows a double set of peaks along the $(h, 0, 0)$
direction at the positions $\bbox{Q} \approx (1 \pm 0.25, 0, 0)$ and $(1 \pm
0.09,0, 0)$. The intrinsic nature of such peaks was demonstrated by
the observation of a signal of similar intensity around the
symmetry-related $(0,1,0)$ point and the presence of four peaks in the
``perpendicular" scan through $(1,0,0)$ shown in
Figure~\ref{fig:qdep}(b). In addition, the variation of intensity
within each set of peaks is quantitatively consistent with the rapid
falloff of the Ru magnetic form factor with the magnitude of
$\bbox{Q}$~\cite{Intl_tables}, providing strong evidence for the
magnetic nature of the excitations.
%
Furthermore, as shown later, the
intensity does not increase with temperature as would be expected if
the scattering were due to lattice vibrations. The extent of the
magnetic fluctuations in the $(h,k,0)$ plane of reciprocal space was
established by the scans shown in Fig.~1(c) and (d). These show that
the excitations give a broad peak centred on the $(h,0,0)$ axis.
The results are summarised in the inset: the excitation
intensity peaks at two incommensurate wavevectors of the form
$\bbox{q}_{\delta} \approx \{0.25,0,0\}$ and
$\bbox{q}_{\epsilon} \approx \{0.09,0,0\}$,
distributed symmetrically about $(1,0,0)$.
It is natural to assume that these arise from peaks in the
wavevector-dependent susceptibility at nesting vectors of the
Sr$_3$Ru$_2$O$_7$ Fermi surface. This is not yet known
experimentally, although it has been calculated~\cite{Hase,Mazin}. The
coupling between the two halves of a bilayer splits each of the three
sheets observed in Sr$_2$RuO$_4$. This, and the rotation of
the octahedra cause hybridisation between the bands. It
appears from the calculations~\cite{Mazin}, that compared with
Sr$_2$RuO$_4$, much of the nesting at the Fermi level is removed,
except between parts of the $\alpha$ sheets (Ru $d_{xz}$ and $d_{yz}$ orbitals).
The calculated sheets have nesting vectors along the (tetragonal)
$\{1,0,0\}$ directions with values which are comparable with (although not equal to)
those we observe. It seems reasonable to conclude that the differences of our
results from those on single-layer Sr$_2$RuO$_4$~\cite{Sidis,Servant}
arise from the effects of bilayers and octahedral rotation
on the Fermi surface in our system.
 %
\begin{figure}
\centerline{\epsfysize=4.6cm \epsfbox{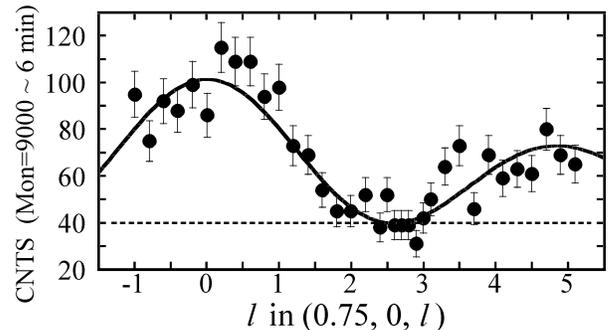}}
\caption{$l$-dependence of inelastic scattering at 2 meV
and $h = 0.75$, showing the effects of the bilayers.
The solid line is the fit described in the text,
plus a constant background. These measurements were performed at a
constant incident energy of 14.67~meV, with a PG higher order filter
in the incident beam.}
\label{fig:elldep}
\end{figure}

Measurements as a function of $l$ allow us to determine the
fundamental fluctuating unit in Sr$_3$Ru$_2$O$_7$ in this energy range.
Fig.~2 shows the variation along $\bbox{c}$* of the intensity
of the signal at $\bbox{q}_{\delta}$.
The experimental data are well-represented by
$I \propto f(\bbox{Q})^2 \cos^2 (2 \pi l z/c)$, where $f(\bbox{Q})$ is the Ru
form factor and $2z =0.194c$ is the distance between the RuO$_2$ planes
in a bilayer.
This function corresponds to the two halves of a bilayer
fluctuating {\it in phase} with each other, but with {\it no} correlation
between bilayers, so that the fluctuations are effectively
two-dimensional. A similar argument~\cite{Sato} was used to
demonstrate 2D fluctuations in YBCO, but
with the two halves of the bilayer in {\it antiphase}.
We point out an important consequence of our results: since
$(1,0,0)$ {\it is} a reciprocal lattice point of a RuO$_2$ bilayer,
the values of the $\bbox{q}$-vectors of excitations should be measured
from this point, rather than $(0,0,0)$ or $(1,0,1)$, which are the closest
reciprocal lattice points of the 3D crystal structure.

\begin{figure}
\centerline{\epsfysize=11.6cm \epsfbox{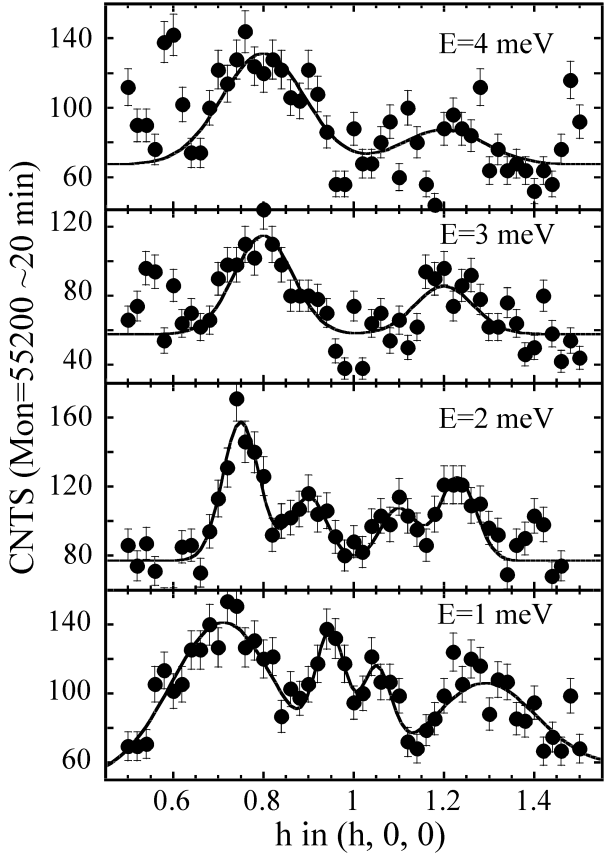}}
\caption{Energy and $\bbox{Q}$-dependence of scattering at 1.5~K.
Scans along $(h,0,0)$ with energy transfers of 1,2,3 and 4~meV.}
\label{fig:endep}
\end{figure}

We now consider the energy dependence of these excitations.
Fig.~\ref{fig:endep} shows four representative $\bbox{Q}$-scans
with energy transfers of 1,2,3 and 4~meV at $T=1.5$~K.
%
The peaks appear to disperse slightly, and merge at higher energies. We have
also performed a $\bbox{Q}$-scan over this region at zero energy
transfer, which showed no evidence for static magnetic ordering
near $\bbox{q}_{\delta}$ or $\bbox{q}_{\epsilon}$. This result is in
agreement with those of~\cite{ICNS,Huang}. We conclude that at
finite temperature only finite frequency, short-range magnetic
correlations exist.

Fig.~4 shows the energy dependence of the
signal at $\bbox{q}_{\delta}$. We have fitted the response to a
simple Lorentzian model for the susceptibility:
$\chi^{\prime \prime}(\bbox{Q},\omega) = \chi^{\prime}(\bbox{Q}) \times
\omega \Gamma(\bbox{Q}) / \left( \Gamma^2(\bbox{Q})+\omega^2 \right)$.
We find that the characteristic energy $\hbar \Gamma = 2.3 \pm 0.3$~meV,
which is much less than 9~meV reported in the single layer
compound Sr$_2$RuO$_4$~\cite{Sidis}. The presence of dispersion on
an energy scale much smaller than $\epsilon_F$ and the small energy scale
of the fluctuations indicates the strong renormalising effects of electron
correlations in our compound. We also note that the susceptibility is
large, translating to $\chi^{\prime}(\bbox{Q}_\delta)$ of
$1.6 \times 10^{-2}$~emu/mol Ru. This indicates that Sr$_3$Ru$_2$O$_7$
is much closer to magnetic order than its sister compound.

We have also followed the fluctuations as a function of temperature
at an energy transfer of 3.1~meV around $\bbox{Q}_{\delta} =
(0.75,0,0)$ (Fig.~5).
At base temperature, the two peaks associated with the
incommensurate spin fluctuations are well defined and
intense. However, as the temperature is increased, the intensity of
the incommensurate peaks falls off, and is replaced by
a broad peak of similar intensity around the 2D reciprocal lattice
point $(1, 0, 0)$.
This position is not a Bragg peak of either the tetragonal or the
orthorhombic cell, so does not give rise to a low energy acoustic
phonon. Hence the peak at $(1, 0, 0)$ is most likely of magnetic origin.
We have confirmed by measurements along $\bbox{c}$* that this
signal also arises from fluctuations of a bilayer unit. Our findings
point to a crossover in the nature of the low energy magnetic correlations
in this material. At high temperatures, 2D ferromagnetic
fluctuations dominate the correlations; as the temperature is
lowered, instead of converging to a long-lived ferromagnetic
state, the system is sidetracked to a different behaviour with
antiferromagnetic finite frequency 2D excitations.
%
\begin{figure}
\centerline{\epsfysize=4.5cm \epsfbox{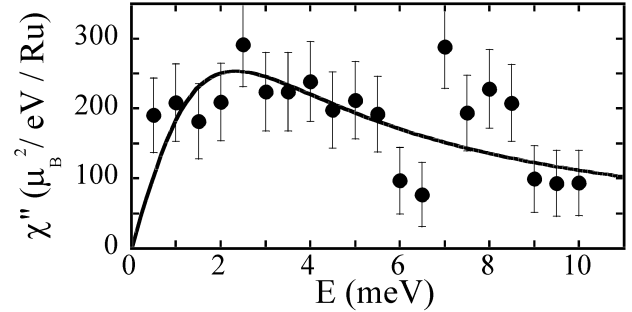}}
\caption{Energy dependence of magnetic scattering at $\bbox{Q} = (0.75,0,0)$,
minus backgrounds taken at $(1.48,0,0)$. The line represents a
Lorentzian as described in the text. The ordinate has been
corrected by the Bose factor, $(n(\omega) + 1 )$ and the Ru form factor and
converted to absolute units (with an accuracy $\sim$20\%)
by normalisation to the intensity of a
transverse acoustic phonon, measured at 3.1~meV around $(1,1,0)$ at
100~K.}
\label{fig:absolute}
\end{figure}
%
\begin{figure}
\centerline{\epsfysize=9.2cm \epsfbox{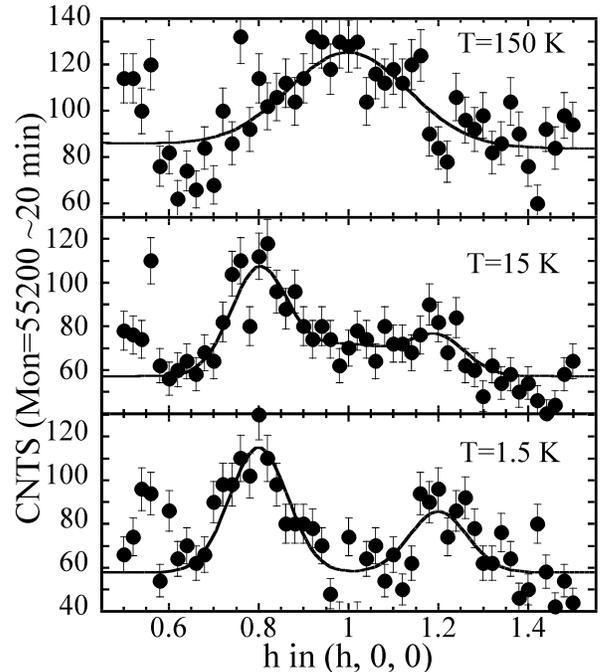}}
\caption{$\bbox{Q}$-dependence of scattering along $(h,0,0)$ for
an energy transfer of 3.1~meV at three different temperatures: 1.5, 15
and 150~K.}
\label{fig:tempdep}
\end{figure}
%
\begin{figure}
\centerline{\epsfysize=11.0cm \epsfbox{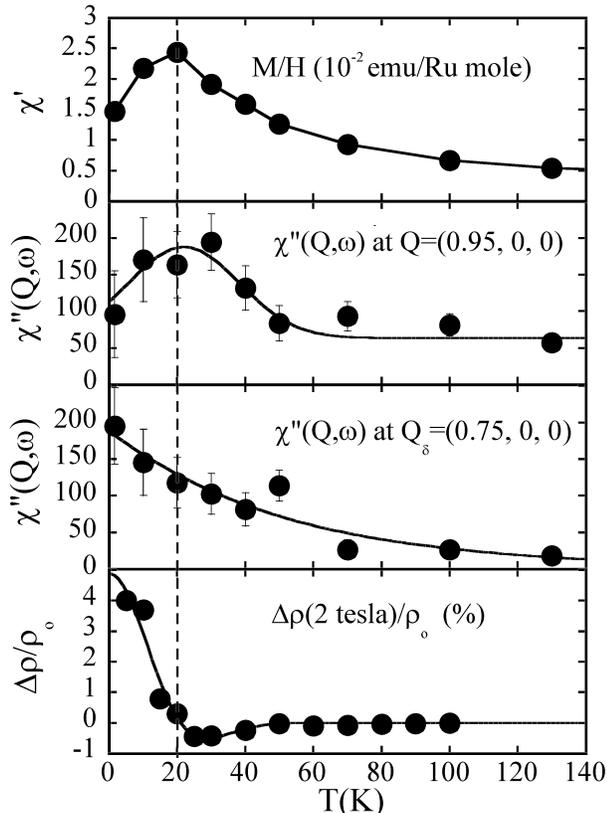}}
\caption{Temperature-dependence of magnetic response from macroscopic
and microscopic measurements:
(a) Static susceptibility from~\protect\cite{Ikeda}.
(b) Susceptibility, $\chi^{\prime \prime}(\bbox{Q},\omega)$
(units: $\mu_{\mbox{B}}^2 / eV / \mbox{Ru}$) from neutron scattering at
$\bbox{Q} = (0.95,0,0)$ and an energy transfer of 2~meV, minus
a background at $(0.55,0,0)$.
(c) As for (b) at $\bbox{Q}_\delta = (0.75,0,0)$.
(d) Fractional magnetoresistance, $(\rho (2T) -\rho (0) )/ \rho (0)$,
measured with current parallel to the magnetic field in
the basal plane~\protect\cite{Rob,Robthesis}.
The lines serve as guides to the eye.}
\label{fig:macro}
\end{figure}

In Fig.~6, we show that the change with temperature in the nature of
magnetic fluctuations is reflected in macroscopic properties. At a
temperature $\approx$20~K, there is a peak in the magnetic
susceptibility~(a), and and also in the susceptibility close to a
ferromagnetic position measured by neutron scattering~(b). The
antiferromagnetic fluctuations~(c) fall away rapidly with
increasing temperature and this is reflected in the change in sign
of the longitudinal magnetoresistance~(d)~\cite{Liu}.
It is not clear what causes this dramatic change in magnetic
correlations, but it may be related to a loss in c-axis electronic
coherence, which reveals itself as a steep rise in the c-axis
resistivity in this temperature region~\cite{Ikeda}.
It is of interest that in the
compound Ca$_{2-x}$Sr$_x$RuO$_4$~\cite{Nakatsuji,Friedt},
doping with Sr drives the system from an insulating antiferromagnetic
state, through a phase with a ferromagnetic instability
to a metallic superconducting one.  In contrast, in
Sr$_3$Ru$_2$O$_7$ the competing interactions coexist in the same
high-quality stoichiometric samples.

In conclusion, we have observed strong 2D spin fluctuations of the
bilayers in Sr$_3$Ru$_2$O$_7$. At high temperatures these
fluctuations are predominantly ferromagnetic in nature, and cross
over to incommensurate ones at low temperatures, with wavevectors
close to those expected for nesting vectors of the Fermi surface.
The small characteristic energy of these fluctuations and their
ambivalent nature suggests that they are implicated in and
related to the metamagnetic transition observed at low
temperatures. We note that a strong temperature-dependence of
the electronic properties and magnetic excitations is also
observed in High T$_c$ superconductors~\cite{Aeppli97}
and heavy fermion systems~\cite{Aeppli88}. Thus the behaviour of
Sr$_3$Ru$_2$O$_7$ may ultimately be related to its proximity to a
quantum critical point~\cite{Grigera}.

\bibliographystyle{prsty}

\noindent
$^{*}$\small{Present address: FKF, Max Planck Institut, Stuttgart, D-70569, Germany;
e-mail: l.capogna@fkf.mpg.de}\\
$^{\dagger}$\small{e-mail: e.m.forgan@bham.ac.uk}\\
$^{\S}$\small{Present address: Electrotechnical Laboratory, Tsukuba, Ibara 305-8568, Japan} \\
\end{document}